\documentclass[preprint,aps,showpacs]{revtex4}
\usepackage{graphicx}

\begin{document}

\title{
Application of Absorbing Boundary Condition 
to Nuclear Breakup Reactions}

\author{M. Ueda}
\email{mueda@nucl.ph.tsukuba.ac.jp}

\author{K. Yabana}
\email{yabana@nucl.ph.tsukuba.ac.jp}

\affiliation{
Institute of Physics, University of Tsukuba, Tsukuba 305-8571, Japan}

\author{T. Nakatsukasa}
\email{takashi@nucl.phys.tohoku.ac.jp}

\affiliation{
Department of Physics, Tohoku University, Sendai 980-8578, Japan}

\date{\today}

\begin{abstract}
Absorbing boundary condition approach to nuclear breakup reactions
is investigated.  A key ingredient of the method is an
absorbing potential outside the physical area,
which simulates the outgoing boundary condition for scattered waves. 
After discretizing the radial variables, the problem results in a linear 
algebraic equation with a sparse coefficient matrix, to which efficient 
iterative methods can be applicable. No virtual state such as 
discretized continuum channel needs to be introduced in the method. 
Basic aspects of the method are discussed by considering a nuclear 
two-body scattering problem described with an optical potential. 
We then apply the method to the breakup reactions of deuterons described
in a three-body direct reaction model. Results employing the absorbing 
boundary condition are found to accurately coincide with those of
the existing method which utilizes discretized continuum channels.
\end{abstract}


\maketitle

\section{Introduction} \label{sec1}

In nuclear reactions with weakly-bound projectiles, breakup processes
are significant not only for their prominent cross sections as a final 
reaction product but also for their influences on other reaction 
channels such as elastic scattering. The breakup processes have 
been investigated originally for the reactions involving weakly-bound 
stable nuclei, such as  deuterons  \cite{yahiro,austern} and $^{6,7}$Li 
\cite{sakuragi}. In the last decade, interests on the breakup 
reactions have explosively increased because of the discovery of the halo 
nuclei around the neutron drip-line and the studies of their reactions 
\cite{tanihata}.

A variety of theories have been developed to describe the breakup 
processes . In this article, we consider reactions in which the 
weakly-bound projectile is described as a composite two-body system
and the whole reaction is described as a three-body problem of 
the projectile and the target. 
At sufficiently high incident energies, one may assume that the projectile 
internal motion is much slower than the projectile-target relative motion. 
Then, the projectile internal motion may be frozen during the reaction.
This is the adiabatic approximation, which has been developed originally
for the deuteron reactions \cite{johnson}. One may also adopt the eikonal 
approximation for the fast projectile-target relative motion. A combined 
use of the eikonal and adiabatic approximations has provided especially 
useful descriptions for the reactions of halo nuclei \cite{YOS}.

For a more accurate description, a fully quantum mechanical treatment of 
the breakup processes has been developed in a coupled channel framework. 
This is known as the coupled discretized continuum channels (CDCC)
method \cite{yahiro,austern,sakuragi}. In this method, the continuum 
excited states of the projectile are incorporated by discretizing the 
momentum variables. The CDCC method has been successful in describing the
reactions of deuterons \cite{yahiro,austern} and $^{6,7}$Li \cite{sakuragi}, 
and has been recently applied to the reactions of halo nuclei \cite{tostevin}.

In the present article, we discuss an alternative, fully quantum mechanical 
method: the method of absorbing boundary condition (ABC). This method was 
developed in the field of chemical reactions \cite{kosloff,neuhasuer,seideman}. The key ingredient is  an introduction of the absorbing potential outside 
the physically relevant area. This absorbing potential allows us to treat 
scattering problems with the vanishing boundary condition. We show
that the ABC method provides a useful description for the breakup 
reactions in which the final scattering channels include three-body 
continuum states.

The present article is organized as follows: In Section II,
we illustrate the ABC method in the simplest example, a two-body 
scattering problem with a nuclear optical potential. In Section III,
we formulate the ABC method for three-body scattering problems.
In Section IV, we present results of the ABC method applied to the breakup
reaction of $d+^{58}$Ni at the incident deuteron energy of 80 MeV.
The results are then compared with those by the CDCC method.
The summary is given in Section V.

\section{Absorbing Boundary Condition method: a potential scattering}
\label{sec2}

We first explain the ABC method considering the simplest example:
the scattering of a particle of mass $m$ from a potential $V({\bf r})$.
We separate the potential $V({\bf r})$ into two parts,
\begin{equation}
V({\bf r}) = V_0({\bf r}) + \Delta V({\bf r}).
\end{equation}
The potential $V_0({\bf r})$ is spherical and may be  of long-range .
We assume that it is easy to construct the scattering solution
$\psi_0^{(+)}({\bf r})$ for the potential $V_0({\bf r})$.
The potential $\Delta V({\bf r})$ may not
be spherical but must be  of short-range , 
$\Delta V({\bf r}) = 0 $ for the spatial region beyond a certain 
radius $r_c$, $\vert {\bf r} \vert > r_c$. 

The scattering wave function can be written in a form
\begin{equation}
\psi^{(+)}({\bf r}) = \psi_0^{(+)}({\bf r})
+ \Delta \psi^{(+)}({\bf r}) \ .
\label{lseq}
\end{equation}
The function $\Delta \psi^{(+)}({\bf r})$ is expressed by
\begin{equation}
\Delta \psi^{(+)}({\bf r}) = 
\int d{\bf r}' G_0^{(+)}({\bf r},{\bf r'}) \Delta V({\bf r'}) 
\psi^{(+)}({\bf r'}) =
\int d{\bf r}' G^{(+)}({\bf r}, {\bf r'}) \Delta V({\bf r'}) 
\psi_0^{(+)}({\bf r'}),
\label{scateq}
\end{equation} 
where $G_0^{(+)}({\bf r},{\bf r'})$ and $G^{(+)}({\bf r},{\bf r'})$ 
are the Green's functions with the potentials $V_0({\bf r})$ and 
$V({\bf r})$, respectively,
\begin{equation}
G_0^{(+)} = 
\frac{1}{E+i\epsilon- T_{\bf r} - V_0} ,
\hspace{1cm}
G^{(+)} = \frac{1}{E+i\epsilon- T_{\bf r}-V}.
\label{greengreen}
\end{equation}
Here, $T_{\bf r}$ is  the  kinetic energy operator. 
A positive infinitesimal 
number $\epsilon$ specifies the outgoing-wave boundary condition.
For the special case of $V_0=0$ with $\Delta V=V$,
$\psi_0^{(+)}({\bf r})$ becomes a plane wave and
we have an analytic expression for $G_0^{(+)}$.

Now, let us explain the ABC method to construct the wave function 
$\Delta \psi^{(+)}({\bf r})$. 
The basic trick is that the positive infinitesimal number $\epsilon$
is replaced by a finite, space-dependent function $\epsilon({\bf r})$.
In order to simulate  the  outgoing boundary condition, 
we introduce the function 
$\epsilon({\bf r})$ which is positive and finite in the spatial region,
$r_c \leq r \leq r_c + \Delta r$. We impose  the  vanishing boundary 
condition at $r= r_c + \Delta r$ for $\Delta \psi^{(+)}({\bf r})$. 
If the absorbing potential works ideally, only  the  outgoing 
waves are allowed in the spatial region around $r = r_c$.
If we regard $\epsilon({\bf r})$ as a part of the Hamiltonian, 
namely ${\cal H} = T_{\bf r}+V-i\epsilon({\bf r})$, one may regard
this replacement as an addition of the absorbing potential,
$-i\epsilon({\bf r})$, to the Hamiltonian, $H = T_{\bf r}+V$. 

Employing the absorbing potential $\epsilon({\bf r})$, we can 
rewrite Eq.~(\ref{scateq}) for $\Delta \psi^{(+)}({\bf r})$ as 
the following linear inhomogeneous equation,
\begin{equation}
(E+i\epsilon({\bf r})-T_{\bf r}-V) \Delta \psi^{(+)}({\bf r}) = 
\Delta V({\bf r}) \psi_0^{(+)}({\bf r}),
\label{shw}
\end{equation}
where, as mentioned above, the scattered wave 
$\Delta \psi^{(+)}({\bf r})$ should satisfy the vanishing boundary
condition at $r=r_c+\Delta r$. The right hand side of this equation
is spatially localized since we assume that the potential 
$\Delta V({\bf r})$ vanishes in the spatial region $r > r_c$. 

The ABC approach thus allows us to obtain  the  scattered wave function
in  the  spatial region of $r < r_c$ by imposing the vanishing boundary 
condition at $r=r_c+\Delta r$. 
We should note that no asymptotic form is required 
to solve the scattering problem in this procedure. The elastic scattering
amplitude $f({\hat {\bf k}})$ can be obtained from 
the wave function in the spatial
region where the potential $\Delta V({\bf r})$ exists,
\begin{equation}
f({\hat {\bf k}}) = f_0({\hat {\bf k}})
-\frac{m}{2\pi\hbar^2} 
\int d{\bf r} \psi_{0{\bf k}}^{(-)*}({\bf r}) \Delta V({\bf r})
\psi^{(+)}({\bf r}),
\end{equation}
where $f_0({\hat {\bf k}})$ is the scattering amplitude with the potential
$V_0({\bf r})$ and $\psi_{0{\bf k}}^{(-)}({\bf r})$ is the incoming-wave
solution in the potential $V_0({\bf r})$ with
the incident momentum ${\bf k}$
($k=\sqrt{2mE}/\hbar$ and ${\hat {\bf k}} = {\bf k}/k$).
Here, the wave functions are normalized as $|\psi({\bf r})|\rightarrow 1$
at $r\rightarrow\infty$.

The accuracy of the calculated solution depends on the quality of the
absorbing potential. One should choose the 
function $\epsilon({\bf r})$ carefully to make the  reflected  waves
as small as possible. In this respect, the absorbing potential
with   linear dependence on  the  coordinate has been tested 
extensively \cite{child91, nakatsukasa}. We parameterize the potential as
\begin{equation}
-i\epsilon({\bf r}) = \left\{
\begin{array}{ccl}
0 & & (r < r_c) \\
 && \\
-i W_{\rm abs} \frac{r-r_c}{\Delta r} & & (r_c \leq r \leq r_c+\Delta r)
\end{array}\right. \ ,
\label{abspot}
\end{equation}
where $W_{\rm abs}$, $r_c$, and $\Delta r$ are positive constants, 
representing the strength,   radius, and   thickness of the 
absorbing potential, respectively. 
The absorbing potential should be sufficiently strong to absorb all 
the outgoing waves, and should be smooth enough to avoid 
the occurrence of the significant reflected waves. 
These conditions could be fulfilled 
if we employ a large enough value for $\Delta r$. The large $\Delta r$, 
 however , implies  the  necessity of treating  a 
large spatial area, which in turn results in the increase of 
the computational task. 
In practice, the following condition derived in the WKB approximation,
gives a criterion for a good absorber \cite{child91, nakatsukasa}:
\begin{equation}
20 \frac{\hbar E^{1/2}}{\Delta r \sqrt{8m}} < W_{\rm abs}
< \frac{1}{10} \Delta r \frac{\sqrt{8m} E^{3/2}}{\hbar} .
\label{Wcond}
\end{equation}
The left inequality of Eq.~(\ref{Wcond}) originates from the condition 
that the absorption is strong enough to suppress any reflection at 
$r=r_c+\Delta r$, while the right inequality originates from the 
condition that the reflection at $r=r_c$ is sufficiently small.
Lower the  bombarding  energy $E$ becomes, the wider 
$\Delta r$ is required to fulfill the condition of Eq.~(\ref{Wcond}). 
This is due to the increase of the de-Broglie wave length 
at low energies.

In order to demonstrate  the  quality of the ABC calculation, we consider,
as an example,  the elastic scattering of 
$^{16}$O + $^{12}$C at $E_{\rm lab.}$ = 139.2 MeV 
with the optical potential of Ref.~\cite{brandan}. The potential $V({\bf r})$ 
is spherical and composed of  the  real and imaginary parts 
of  the  nuclear potential and of the Coulomb potential whose short range 
part is regularized with a quadratic function 
(the Coulomb potential of a uniformly charged sphere).

Although one can easily obtain the numerical solution for the potential
$V({\bf r})$, we divide the potential into two parts, $V_0(r)$ and
$\Delta V(r)$, in order to test the ABC. Here, we adopt the real part 
of the nuclear potential as $\Delta V(r)$, and the rest as $V_0(r)$.
We solve Eq.~(\ref{shw}) in the partial wave expansion,
$\Delta \psi^{(+)}({\bf r}) = \sum_{lm} (y_l(r)/r) Y_{lm}({\hat {\bf r}})$.

We fix the parameters of the absorbing potential, Eq.~(\ref{abspot}),
as $r_c=20$ fm and $\Delta r=10$ fm, but vary $W_{\rm abs}$. 
The vanishing boundary condition, $y_l(r)=0$, is imposed at 
$r=r_c+\Delta r = 30$ fm. The discrete variable representation,
which will be explained in detail in Section III, is utilized
with the uniform grid spacing of 0.2 fm.

\begin{figure*}
\includegraphics[width=100mm]{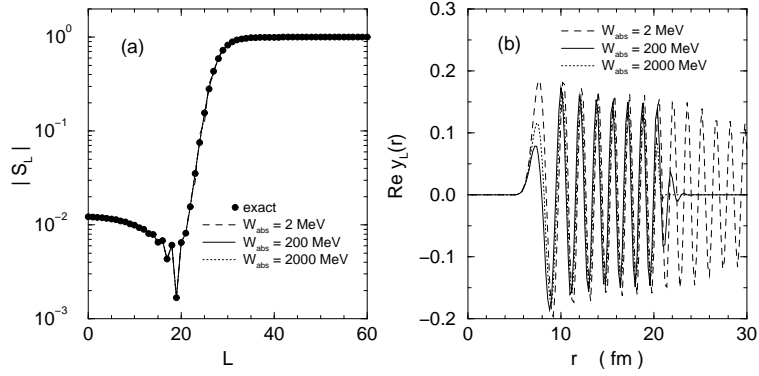}
\caption{\label{fig1}
(a) The absolute values of the $S$-matrices as a function of 
the relative angular momentum $l$ for the elastic $^{16}$O + $^{12}$C 
scattering at $E_{\rm Lab.}$ = 139.2 MeV. (b) The real part of the 
radial wave function of the scattering partial wave $l = 30$.}
\end{figure*}

In Fig.~\ref{fig1} (a), we show
absolute values of  the  $S$-matrices as a 
function of $l$ calculated with different  values of $W_{\rm abs}$ . 
The closed circles represent  the  absolute values of the $S$ matrices 
which are obtained by a standard numerical procedure to integrate 
the radial Schr\"odinger equation up to $r$ = 30 fm. One can see that 
the ABC method works quite well in describing the potential 
scattering. It should be noted 
that only the absorbing potential with $W_{\rm abs}$ = 200 MeV 
satisfies the condition of Eq. (\ref{Wcond}). 

In Fig. \ref{fig1} (b), we show the real part of $y_l(r)$ for 
the partial wave of $l = 30$, which corresponds to the grazing angular 
momentum. The three lines correspond to  the  different choices of the
strength  $W_{\rm abs}$.
The damping behavior of the three wave functions 
in the region $20 \le r \le 30$ fm are different according to the strength 
$W_{\rm abs}$. Despite of small deviation in the wave functions
at $r < 20$ fm, it does not affect the absolute 
values of the elastic scattering $S$ matrix elements as we saw
above. This implies that the reaction cross section is not so sensitive
to the choice of the absorbing potential. 

\begin{figure*}
\includegraphics[width=100mm]{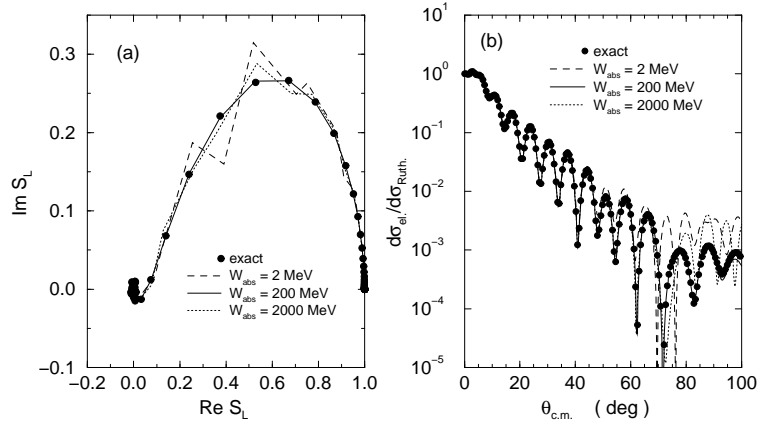}
\caption{\label{fig2}
The elastic scattering $S$ matrix elements $S_l$ (a) and the
differential cross section (b) for the same reactions as shown in 
Fig.~\ref{fig1}.}
\end{figure*}

However, when evaluating  the  differential cross sections,
one has to be careful on the choice of the absorbing potential. 
Figure~\ref{fig2} displays the Argand 
diagram of the elastic scattering
$S$ matrix elements $(a)$, and the differential cross sections $(b)$. 
The elastic scattering cross sections in the Rutherford ratio
vary by a few orders of magnitude as a function of the
scattering angle.  The  subtle difference in the scattering matrix
results in  the  substantial deviation in the differential cross sections. 
The calculation with $W_{\rm abs}$=200 MeV, which fulfills the condition
of Eq.~(\ref{Wcond}), is reasonably accurate while the other choices 
of $W_{\rm abs}$, which do not fulfill the condition, give less accurate 
results at large angles.

\section{Absorbing Boundary Condition method: a three-body reaction}
\label{sec3}

\subsection{Basic equations}

In this section we explain how to apply the ABC to three-body 
breakup reactions. We consider the reaction of a projectile ($P$) on 
a target ($T$), in which the projectile is composed of a core 
nucleus ($C$) and a neutron ($n$). 
We denote masses of the core, the neutron, 
and the target as $m_C$, $m_n$, and $m_T$, respectively. 
The charge numbers of the core and   target are
$Z_C$ and $Z_T$, respectively.
We employ a Jacobi coordinate system in which the $n$-$C$ relative 
coordinate is denoted as ${\bf r}$ and  the  $P$-$T$ relative coordinate 
as ${\bf R}$. 
We assume that the total Hamiltonian is given by the following form,
\begin{equation}
H = -\frac{ \hbar^2 }{2\mu} {\bf \nabla}_{\bf R}^2 
-\frac{ \hbar^2 }{2m} {\bf \nabla}_{\bf r}^2 
+ V_{nC}({\bf r}_{nC}) 
+ V_{nT}({\bf r}_{nT}) + V_{CT}({\bf r}_{CT}) ,
\label{th} 
\end{equation}
where the reduced masses, $\mu$ and $m$, of the relative $P$-$T$ and 
$n$-$C$ motions, respectively, are given by 
\begin{equation}
\mu = \frac{(m_n + m_C)m_T}{(m_n+m_C)+m_T} \ , \hspace{1cm}
m = \frac{m_n m_C}{m_n+m_C} .
\end{equation}
The vectors 
\begin{equation}
{\bf r}_{nC} = {\bf r} \ , \hspace{1cm}
{\bf r}_{nT} = {\bf R}+\frac{m_C}{m_n+m_C} {\bf r} \ ,
\hspace{1cm}
{\bf r}_{CT} = {\bf R}-\frac{m_n}{m_n+m_C} {\bf r} ,
\end{equation}
represent the $n$-$C$, $n$-$T$, and $C$-$T$
separations, respectively. $V_{nC}$, $V_{nT}$, $V_{CT}$ are 
the effective potentials between the corresponding particles. 
$V_{CT}$ includes the Coulomb potential as well as the nuclear potential. 
$V_{nT}$ and $V_{CT}$ may be complex. Their imaginary parts represent 
excitations of the target and/or core nucleus. 
$V_{nC}$ is assumed to be real.

We introduce an auxiliary distorting potential $U_{PT}({\bf R})$ for 
the $P$-$T$ relative motion. The $U_{PT}({\bf R})$ may be chosen
arbitrarily as long as it contains
the long-range Coulomb part of the $P$-$T$ interaction. 
The distorted wave function 
$\Phi^{D(+)}_{\bf K}({\bf R})$ is a scattering solution of the 
following equation with the outgoing boundary condition,
\begin{equation}
\biggl [ -\frac{\hbar^2}{2\mu} {\bf \nabla}_{\bf R}^2 + U_{PT}({\bf R}) 
\biggr ] \Phi^{D(+)}_{\bf K}({\bf R}) = E \Phi^{D(+)}_{\bf K}({\bf R}) \ ,
\hspace{1cm} \biggl ( \, E = \frac{\hbar^2 {\bf K}^2}{2\mu}
\, \biggr )
\end{equation}
where ${\bf K}$ is the incident wave number vector. 
In the partial wave expansion, we express 
 $\Phi^{D(+)}_{\bf K}({\bf R})$ as 
\begin{equation}
\Phi^{D(+)}_{\bf K}({\bf R})
=
4 \pi \sum_{LM} i^L e^{i(\sigma_L + \delta_L)}
\frac{f_L(K,R)}{KR} Y_{LM}^{\ast}({\hat {\bf K}}) Y_{LM}({\hat {\bf R}}).
\end{equation}

For  the  $n$-$C$ motion described with the Hamiltonian
$h_{nC} = -\frac{\hbar^2}{2m}\nabla_{\bf r}^2 + V_{nC}({\bf r})$, 
we denote  the  solutions as $\phi_{\beta}({\bf r})$ , 
where $\beta$  specifies   the  quantum numbers 
$n_{\beta} l_{\beta} m_{\beta}$ for  the  bound orbitals and 
$k_{\beta} l_{\beta} m_{\beta}$ for  the  continuum states.
$n_{\beta}$ is  the  nodal quantum number and $k_{\beta}$
is the magnitude of the wave number. $l_{\beta}$ and $m_{\beta}$ are
the orbital angular momentum quantum numbers as usual.  
In the polar coordinate system,  $\phi_{\beta}({\bf r})$ is rewritten 
as
\begin{equation}
\label{phi_beta}
\phi_{\beta}({\bf r}) = \frac{u_{\beta}(r)}{r} 
Y_{l_{\beta}m_{\beta}}({\hat {\bf r}}),
\end{equation}
where $u_{\beta}=u_{n_{\beta}l_{\beta}}$ for  the  bound orbitals and
$u_{\beta}=u_{k_{\beta}l_{\beta}}$ for  the  continuum  states .
The bound wave function is normalized as usual, and the continuum wave 
function is normalized by the following asymptotic form,
\begin{equation}
u_{k_{\beta}l_{\beta}}(r) \sim \sqrt{\frac{2}{\pi}}
\sin \left( k_{\beta}r - \frac{l_{\beta}}{2}\pi 
+ \delta_{l_{\beta}} \right),
\quad\mbox{at } r\rightarrow\infty,
\end{equation}
where $\delta_{l_{\beta}}$ is  the  phase shift 
 for the  $n$-$C$ scattering.

We express the total wave function 
$\Psi^{(+)}_{{\bf K}_0 \alpha}({\bf R}, {\bf r})$ as  the  
sum of  the  distorted wave in the incident channel $\alpha$ 
and a scattered wave as
\begin{equation}
\Psi^{(+)}_{{\bf K}_0 \alpha}({\bf R},{\bf r})
= \Phi^{D(+)}_{\bf K_0}
({\bf R}) \, \phi_{\alpha}({\bf r}) + \Psi_{\rm S}({\bf R},{\bf r}),
\end{equation}
where ${\bf K}_0$ is  the  incident wave number vector of the $P$-$T$
relative motion.

The scattered wave funtion $\Psi_{\rm S}({\bf R},{\bf r})$,
satisfies the following equation
\begin{equation}
\left( E  -H \right)
\Psi_{\rm S}({\bf R},{\bf r})  
= \Delta V({\bf R}, {\bf r})
\, \Phi^{D(+)}_{\bf K_0}({\bf R}) \, \phi_{\alpha}({\bf r}) ,
\label{3Bscat} 
\end{equation}
with
\begin{equation}
\Delta V({\bf R}, {\bf r}) = V_{nT}({\bf r}_{nT})+V_{CT}({\bf r}_{CT}) 
- U_{PT}({\bf R}) \ ,
\end{equation}
where $E$ is the total energy which is equal to the sum of the
energies of the incident $P$-$T$ relative motion 
$\hbar^2 {\bf K}_0^2/2\mu$ and the energy of  the 
$n$-$C$ relative motion in the incident channel, $\epsilon_{\alpha}$.
The right hand side,
$\Delta V({\bf R}, {\bf r}) \Phi^{D(+)}_{\bf K_0}({\bf R}) 
\phi_{\alpha}({\bf r})$, 
should be localized in space. In other words, this function vanishes
if either $R$ or $r$ is larger than a certain critical radius,
$R_C$ for $R$ and $r_c$ for $r$ \footnote{
The Coulomb potential could induce a long-range tail of
$\Delta V(R,r)$  because of the difference between the center of
mass and the center of charge for a projectile.
An extended spatial region is required if the breakup reactions
induced by this Coulomb field are significant.}.

Since the scattered wave function, $\Psi_{\rm S}({\bf R},{\bf r})$,
includes only  the  outgoing waves asymptotically, 
it should be legitimate to simulate
the outgoing boundary condition by introducing the 
absorbing potentials for both $R$ and $r$ coordinates.
Thus, instead of solving Eq.~(\ref{3Bscat}) with the outgoing boundary
condition, we propose to employ the following equation with the vanishing
boundary condition.
\begin{equation}
\left[ E + i\epsilon_{nC}(r) + i\epsilon_{PT}(R) -H \right]
\Psi_{\rm S}({\bf R},{\bf r})  
= \Delta V({\bf R}, {\bf r})
\, \Phi^{D(+)}_{\bf K_0}({\bf R}) \, \phi_{\alpha}({\bf r}) ,
\label{3Babs} 
\end{equation}
where $\epsilon_{nC}(r)$ and $\epsilon_{PT}(R)$ are  the  absorbing
potentials for the relative motions of $n$-$C$ and $P$-$T$,
respectively. They are placed in the spatial region
$R > R_C$ for $\epsilon_{PT}(R)$, and $r > r_C$ for
$\epsilon_{nC}(r)$. In practice, we employ the linear absorbing 
potentials of Eq.~(\ref{abspot}) for both functions. The
vanishing boundary condition is imposed for $\Psi_{\rm S}({\bf R},{\bf r})$ at
$R=R_C+\Delta R$ and $r=r_C+\Delta r$.
In this way, the three-body scattering problem is converted
into  the  three-body Schr\"odinger-like equation with 
 the  vanishing 
boundary condition. No explicit consideration on the boundary
conditions for three-body continuum states is required in this approach.

As for the parameters of the absorbing potentials, careful
considerations are required.
For $i\epsilon_{PT}(R)$, we choose  the potential parameters 
according to the condition of Eq.~(\ref{Wcond}) , 
where the energy $E$ may be evaluated at the relative
$P$-$T$ energy in the incident channel. 
However, for $i\epsilon_{nC}(r)$, the energy of $n$-$C$ motion
after reaction is not unique but spreads over a certain energy region.
A reasonable choice is then to set the parameters of $i\epsilon_{nC}(r)$
optimum for the main component of the $n$-$C$ motion. It should, 
however, be noted that the complete absorption of the breakup waves  
is difficult for small $n$-$C$ relative energies  
because of their long  wavelengths.

The three-body wave function obtained in the above procedure
is meaningful in the inner spatial region, $r < r_C$ and $R < R_C$. 
The wave function only in
this interacting region is required to calculate the relevant $T$-matrices.
For the breakup processes in which the final $n$-$C$ motion is specified 
by the relative wave number ${\bf k}$ and the final $P$-$T$ motion 
specified by ${\bf K}$, the relevant $T$-matrices are given by
\begin{equation}
T_{\alpha}({\bf K},{\bf k})
= \langle \Phi_{\bf K}^{D(-)}({\bf R}) \phi_{\bf k}^{(-)}({\bf r}) 
\vert V_{nT}+V_{CT}-U_{PT} \vert
\Psi_{{\bf K}_0 \alpha}^{(+)}({\bf R},{\bf r}) \rangle \ ,
\label{tmat1}
\end{equation}
where $\phi_{\bf k}^{(-)}({\bf r})$ is the $n$-$C$ wave function
with the incident plane wave specified by the wave number vector
${\bf k}$ and with the incoming boundary condition. This is
related to $\phi_{klm}({\bf r})$ of Eq. (\ref{phi_beta}) by
\begin{equation}
\phi_{\bf k}^{(-)}({\bf r}) 
=\sum_{lm} \frac{(2\pi)^{3/2}}{k} i^l e^{-i\delta_l}
\phi_{klm}({\bf r}) Y_{lm}^*({\hat {\bf k}}) \ ,
\end{equation}
where $\delta_l$ is the  phase shift for the $n$-$C$ scattering.
For convenience, we introduce an alternative definition for the
$T$-matrices, $T_{\alpha'\alpha}({\bf K})$, in which 
$\alpha'$ specifies the final $n-C$ state, $\alpha'=nlm$ for the
bound orbitals and $\alpha'=klm$ for the continuum states.
This $T$-matrix is given by
\begin{equation}
T_{\alpha' \alpha}({\bf K})=
\langle \Phi_{\bf K}^{D(-)}({\bf R}) \phi_{\alpha'}({\bf r}) 
\vert V_{nT}+V_{CT}-U_{PT} \vert
\Psi_{{\bf K}_0 \alpha}^{(+)}({\bf R},{\bf r}) \rangle.
\label{tmat2}
\end{equation}
For the reactions in which the final state is the three-body 
continuum state specified by $\alpha' = klm$, the $T$-matrices of
Eqs. (\ref{tmat1}) and (\ref{tmat2})
are related by
\begin{equation}
T_{\alpha}({\bf K},{\bf k})
= \frac{(2\pi)^{3/2}}{k}\sum_{lm} (-i)^l
e^{i\delta_l} Y_{lm}({\hat {\bf k}}) T_{klm, \alpha}({\bf K}).
\end{equation}

Various cross sections can be expressed in terms of these $T$-matrices.
For example, the angle-integrated elastic breakup cross section is given by
\begin{equation}
\sigma^{\rm breakup} = \left( \frac{\mu}{2\pi \hbar^2} \right)^2
\sum_{lm} \int_0^{\infty} dk \frac{K}{K_0}
\vert T_{klm,\alpha}({\bf K}) \vert^2 \ .
\end{equation}

\subsection{Radial coupled channel equation}
\label{subsec31} 

In practice, Eq.~(\ref{3Babs}) is solved in the partial wave expansion.
The incident wave function 
$\Phi_{{\bf K}_0}^{(+)}({\bf R}) \phi_{\alpha}({\bf r})$
 and  the total wave function $\Psi^{(+)}_{{\bf K}_0 \alpha}
({\bf R}, {\bf r})$ are expressed as  
\begin{eqnarray}
\Phi_{{\bf K}_0}^{(+)}({\bf R}) \phi_{\alpha}({\bf r})
&=&
\sum_{JL}\sqrt{4\pi(2L+1)}
\langle L 0 l_{\alpha} m_{\alpha} \vert J m_{\alpha} \rangle
i^L e^{i(\sigma_L + \delta_L)} \nonumber \\
&\times& \frac{f_L(K_0,R)}{K_0 R} \frac{u_{n_{\alpha}l_{\alpha}}(r)}{r}
[\, Y_L({\hat {\bf R}}) \, 
Y_{l_{\alpha}}({\hat {\bf r}})\, ]_{J m_{\alpha}},
\label{eq321}
\end{eqnarray}
 and  
\begin{eqnarray}
\Psi^{(+)}_{{\bf K}_0 \alpha}({\bf R},{\bf r}) &=&
\sum_{JLL'l'} \sqrt{4\pi(2L+1)}
\langle L 0 l_{\alpha} m_{\alpha} \vert J m_{\alpha} \rangle
i^L e^{i(\sigma_L+\delta_L)} \nonumber \\
&\times& \frac{y^J_{Ll_{\alpha},L'l'}(R,r)}{Rr}
[Y_{L'}({\hat {\bf R}}) Y_{l'}({\hat {\bf r}})]_{J m_{\alpha}} \ ,
\label{eq322}
\end{eqnarray}
 respectively .  For the scattered wave function 
$\Psi_S^{(+)}({\bf R},{\bf r})$,  we make the same expansion as
Eq.~(\ref{eq322}), replacing the radial wave function 
$y^J_{Ll_{\alpha},L'l'}(R,r)$ with $\alpha^J_{Ll_{\alpha},L'l'}(R,r)$. 
These two radial wave functions,
$y^J_{Ll_{\alpha},L'l'}(R,r)$ and $\alpha^J_{Ll_{\alpha},L'l'}(R,r)$,
are related to each other by
\begin{equation}
y^J_{Ll_{\alpha},L'l'}(R,r)
=
\frac{1}{K_0}f_L(K_0,R)u_{n_{\alpha}l_{\alpha}}(r)
\delta_{LL'} \delta_{l_{\alpha}l'} +
\alpha^J_{Ll_{\alpha},L'l'}(R,r).
\end{equation}

The radial coupled-channel equation for the scattered wave
function $\alpha^J_{Ll_{\alpha,L'l'}}(R,r)$ is  written  as
\begin{eqnarray}
&&\Biggl[ E + i\epsilon_{nC}(r) + i\epsilon_{PT}(R) 
\nonumber\\
&&- \left\{
-\frac{\hbar^2}{2\mu} \frac{\partial^2}{\partial R^2}
+ \frac{\hbar^2 L'(L'+1)}{2\mu R^2}
-\frac{\hbar^2}{2m} \frac{\partial}{\partial r^2}
+ \frac{\hbar^2 l'(l'+1)}{2mr^2} + V_{nC}(r) \right\} \Biggr]
\alpha^J_{Ll_{\alpha}, L'l'}(R,r) \nonumber\\
&&-
\sum_{L''l''}V^J_{L'l',L''l''}(R,r)
\alpha^J_{Ll_{\alpha},L''l''}(R,r) \nonumber\\
&=&
\left\{ V^J_{L'l',Ll_{\alpha}}(R,r)
-\delta_{LL'}\delta_{l_{\alpha}l'}U_{PT}(R) \right\}
\frac{1}{K_0} f_L(K_0,R)u_{n_{\alpha}l_{\alpha}}(r) \ ,
\label{3Brad}
\end{eqnarray}
where the radial coupling potential
$V^J_{L'l',L''l''}(R,r)$
is defined by
\begin{equation}
V^J_{L'l',L''l''}(R,r)
= \int d{\hat {\bf R}} d{\hat {\bf r}}
[Y_{L'}({\hat {\bf R}})Y_{l'}({\hat {\bf r}})]^*_{JM} (V_{nT}+V_{CT})
[Y_{L''}({\hat {\bf R}}) Y_{l''}({\hat {\bf r}})]_{JM} \ .
\end{equation}

The $T$-matrix elements are expressed in terms of the radial wave
functions  . For a transition  concerning the $n$-$C$ relative
motion  from  the bound orbital
$\alpha=n_{\alpha} l_{\alpha} m_{\alpha}$ 
 in the incident channel  to the continuum state 
$\alpha' = k l m$  in the final channel, the relevant $T$-matrix
element is expressed as
\begin{eqnarray}
T_{\alpha' \alpha}({\bf K})
&=&
\langle \Phi_{\bf K}^{D(-)}({\bf R}) \phi_{\alpha'}({\bf r})
\vert V_{nT} + V_{CT} - U_{PT} \vert
\Psi_{{\bf K}_0 \alpha}^{(+)}({\bf R},{\bf r}) \rangle
\nonumber\\
&=&
\sum_{JLL'} (4\pi)^{3/2}(2L+1)^{1/2}i^{L-L'}
e^{i\sigma_L+i\sigma_{L'}+i\delta_L+i\delta_{L'}}
\nonumber\\
&&\times
\langle L 0 l_{\alpha} m_{\alpha} \vert J m_{\alpha} \rangle
\langle L' m_{\alpha}-m l m \vert J m_{\alpha} \rangle
Y_{L' m_{\alpha}-m}({\hat {\bf K}}) I^J_{\alpha' L', \alpha L} \ ,
\label{eq30}
\end{eqnarray}
where the radial integral
$I^J_{\alpha' L', \alpha L}$
is given by
\begin{equation}
I^J_{\alpha' L', \alpha L}
=
\sum_{L''l''} \int_0^{\infty} dR dr \frac{1}{K} f_{L'}(K,R) u_{\alpha'}(r)
\left\{ V^J_{L'l,L''l''}(R,r)-\delta_{L'L''}\delta_{ll''}U_{PT}(R) \right\}
y^J_{Ll_{\alpha},L''l''}(R,r).
\end{equation}
For the elastic channel in which the final state $\alpha'$ is identical 
to the initial one, $\alpha'=\alpha=n_{\alpha}l_{\alpha}m_{\alpha}$, the
two-body $T$-matrix for $U_{PT}$ should be added to the above expression of
Eq.~(\ref{eq30}).

For later convenience, we present an expression for the angle-integrated
elastic breakup cross section in terms of this radial integral.
The elastic breakup cross section can be decomposed into  the  
sum over the total angular momentum $J$ and the relative angular 
momentum $l$ of the $n$-$C$ motion, as well as  the integral over 
the relative wave number $k$,
\begin{equation}
\sigma^{\rm breakup}
= \sum_{JL'l} \int_0^{\infty} dk \sigma^J_{L'l}(k),
\end{equation}
with
\begin{equation}
\sigma^J_{L'l}(k) = 16\pi \left( \frac{\mu}{\hbar^2} \right)^2
\frac{K}{K_0}
\left| \sum_L  (2L+1)^{\frac{1}{2}}e^{i\delta_L+i\delta_{L'}}
\langle L 0 l_{\alpha} m_{\alpha} \vert J m_{\alpha} \rangle
I^J_{\alpha' L', \alpha L} \right|^2,
\label{sgmJLl}
\end{equation}
where $K$ is the $P$-$T$ relative wave number which implicitly
depends on the $n$-$C$ wave number $k$ through  the 
energy conservation relation.

\subsection{Computational details}

To solve the radial coupled-channel equation (\ref{3Brad}), 
we discretize the radial variable with a uniform grid. 
To achieve high accuracy with a moderate number of grid points, 
we employ  the  so-called discrete variable 
representation (DVR) \cite{DVR}. 
Although the DVR is a kind of basis expansion methods, 
the resultant equation is similar to that in the finite 
difference approximation.

We employ  labels  $i$  and  $j$ for  the  
grid points of  the radial coordinates $R$ and $r$, respectively.
Denoting the grid  spacings  for  the  coordinates $R$ and $r$
as $H$ and $h$, respectively,
the grid points $R_i$ and $r_j$ are given by
\begin{equation}
R_i = i \, H\ , (i=1 \cdots N_R) \hspace{1cm} r_j = j \, h \ , 
(j=1 \cdots N_r) \ ,
\end{equation}
where $R_c+\Delta R = N_R \cdot H$ and $r_c+\Delta r = N_r \cdot h$.
The wave functions are then discretized as
\begin{equation}
[\alpha^J_{Ll_{\alpha},L'l'} ]_{ij}
= \alpha^J_{Ll_{\alpha},L'l'}(R_i,r_j).
\end{equation}
In the DVR, the matrix elements of the potentials become diagonal,
\begin{equation}
[V^J_{Ll,L'l'}]_{ij,i'j'}
= V^J_{Ll,L'l'}(R_i,r_j) \delta_{ii'} \delta_{jj'} \ .
\end{equation}
The kinetic energy operator for the coordinate $R$ is expressed as
the following matrix form,
\begin{equation}
\left[ -\frac{\hbar^2}{2 \mu} \frac{d^2}{dR^2} \right]_{ij,i'j'} 
= \delta_{jj'} \frac{\hbar^2}{2\mu} D^R_{ii'},
\end{equation}
where $D^R_{ii'}$ is defined by
\begin{equation}
D^R_{ii'}
= \frac{(-1)^{i-i'}}{ H^2} \times
\left \{
\begin{array}{lcc} 
\frac{\pi^2}{3}-\frac{1}{2 i^2} & {\rm for} & (i=i') \\
    & & \\
\frac{2}{(i-i')^2}-\frac{2}{(i+i')^2} & {\rm for} & (i\not=i') 
\end{array}
\right . \ \ .
\label{tgrid}
\end{equation}
The kinetic energy operator for the radial coordinate $r$ is
expressed in a similar manner.

After these procedures, the radial coupled-channel equation,
Eq.~(\ref{3Brad}), is expressed as the following matrix form,
\begin{equation}
\sum_{L''l'' i'j'} [A^{J}_{L'l', L''l''}]_{ij,i'j'}
[\alpha^J_{Ll_{\alpha},L''l''}]_{i'j'}
= [b_{L l_{\alpha},L'l'}]_{ ij} \ ,
\end{equation}
where the coefficient matrix $[A^J_{L'l' , L''l''}]_{ij, i'j'}$
is defined by
\begin{eqnarray}
[A^J_{L'l', L''l''}]_{ij,i'j'}
&=& E\delta_{L'L''}\delta_{l'l''}
\delta_{ii'} \delta_{jj'}
\nonumber\\
&&
-\left\{ \frac{\hbar^2}{2\mu}D^R_{ii'} 
+ \left( \frac{\hbar^2 L'(L'+1)}{2\mu R_i^2}
- i\epsilon_{PT}(R_i)\right)\delta_{ii'} \right\}\delta_{jj'} 
\delta_{L'L''}\delta_{l'l''}
\nonumber\\
&& 
-\left\{ \frac{\hbar^2}{2m}D^r_{jj'}
+ \left( \frac{\hbar^2 l'(l'+1)}{2mr_j^2} + V_{nC}(r_j)
-  i\epsilon_{nC}(r_j)\right)\delta_{jj'} \right\} \delta_{ii'}
\delta_{L'L''}\delta_{l'l''}
\nonumber\\
&&
- V^J_{L'l', L''l''}(R_i, r_j)
\delta_{ii'} \delta_{jj'}
\end{eqnarray}
and the source term $[b_{Ll_{\alpha},L'l'}]_{ij}$ is given by
\begin{equation}
[b_{L l_{\alpha},L'l'}]_{ij} =
\left\{ V^J_{L'l', Ll_{\alpha}}(R_i, r_j)
-\delta_{LL'} \delta_{ll_{\alpha}} U_{PT}(R_i) \right\}
\frac{1}{K_0}
f_L(K_0,R_i) u_{n_{\alpha}l_{\alpha}}(r_j) \ .
\end{equation}

The dimension $N_{\rm dim}$ of this linear algebraic equation is given by
 the  product of the number of angular momentum channels , 
which  is denoted by  $N_J$, the  numbers  of grid points 
$N_R$ and $N_r$ , i.e.,  $N_{\rm dim}=N_J N_R N_r$. 
As will be seen in  a  practical example given
in the next section, this dimension becomes as large as $N_{\rm dim}=10^5$.
The coefficient matrix $[A^J_{L'l', L''l''}]_{ij,i'j'}$ is sparse, 
namely, the number of non-zero elements of this matrix is rather small. 

There are many efficient algorithms which are proposed to solve such 
a linear algebraic equation with a sparse coefficient matrix. 
We will employ the Bi-conjugate gradient method \cite{BGM}, 
which is one of the most well-known iterative method to solve the linear
algebraic equation with a complex non-hermitian coefficient matrix. 
In order to improve 
the convergence, the pre-conditioning \cite{precon} is carried out. 
In our problem, the diagonal part of the matrix ${\bf A}$ includes 
huge elements because of the centrifugal barrier. 
 In order to  balance the matrix, we modify the equation as
\begin{equation}
\frac{1}{\sqrt{\bf D}}{\bf A}\frac{1}{\sqrt{\bf D}} \cdot
\sqrt{\bf D}\vec \alpha =
\frac{1}{\sqrt{\bf D}} \vec b \ ,
\end{equation}
where ${\bf D}$ is  the  diagonal matrix whose elements
are equal to the diagonal matrix elements of ${\bf A}$.
The convergence of the Bi-conjugate gradient method is significantly
accelerated for the matrix
$\frac{1}{\sqrt{\bf D}}{\bf A}\frac{1}{\sqrt{\bf D}}$
compared with the original matrix ${\bf A}$.

After solving the equation, 
 we calculate  the $T$-matrices  
with the similar discretization.

\section{Deuteron reactions: Comparison with the CDCC method}
\label{sec4}

We here apply  the ABC method to the $d$+$^{58}$Ni reaction 
at the incident deuteron energy $E_d = 80$ MeV. Since this 
reaction was studied in detail with the CDCC method \cite{yahiro1}, 
we can assess validity of the ABC method by comparing our results 
with those of Ref.~\cite{yahiro1}.

We employ the same Hamiltonian as   used in Ref.~\cite{yahiro1}.
Namely, the global optical potential for nucleon-nucleus elastic 
scattering \cite{becchetti} is adopted as the interactions between 
the proton (neutron) in the deuteron and $^{58}$Ni. The potentials
at half of the incident deuteron energy is adopted. No spin-orbit 
 force is  taken into account. The potential parameters are:
$V_R$ = 44.921 MeV, $r_R$ = 1.17 fm, $a_R$ = 0.75 fm, 
$W_V$ = 6.100 MeV, $W_{SF}$ = 2.214 MeV,  
$r_I$ = 1.32 fm, $a_I$ = 0.534 fm for proton and
$V_R$ = 42.627 MeV, $r_R$ = 1.17 fm, $a_R$ = 0.75 fm, 
$W_V$ = 7.240 MeV, $W_{SF}$ = 2.586 MeV,  
$r_I$ = 1.26 fm, $a_I$ = 0.534 fm for neutron.
The Coulomb potential of uniformly charged sphere is assumed between
the centers of mass of the deuteron and $^{58}$Ni. The radius of the
sphere is set as 4.92 fm.

The potential for the $p$-$n$ relative motion is taken as
a Gaussian form,
$V_{np}(r)=V_0 \exp [-r^2/a_0^2]$ with $V_0=-$72.15 MeV
and $a_0=$ 1.484 fm. The ground state is assumed
to be a pure $s$-wave. This potential gives the deuteron binding 
energy of 2.22 MeV.

We treat the radial regions of $R$ and $r$ up to 50 fm.
The radial scattered wave function $\alpha^J_{Ll_{\alpha},L'l'}(R,r)$
is set equal to zero if either $R$ or $r$ is  larger  than 50 fm.
The absorbing potentials $i\epsilon_{PT}(R)$ and $i\epsilon_{nC}(r)$
are active at $R,r >  25$ fm. They are parameterized as
\begin{equation}
i\epsilon_{PT}(R) = {i W_{PT}}\frac{R-R_c}{\Delta R},
\hspace{5mm}
i\epsilon_{nC}(r) = {i W_{nC}}\frac{r-r_c}{\Delta r} \ ,
\end{equation}
with
$W_{PT}$ = 50 MeV, $R_c$ = 25 fm, $\Delta R$ = 25 fm,
$W_{nC}$ = 20 MeV, $r_c$ = 25 fm, and $\Delta r$ = 25 fm.
This implies that the obtained solution will be reliable
in the spatial region of $R,r < 25$ fm.
As for the auxiliary distorting potential $U_{PT}({\bf R})$,
we take the folding
potential in which the $p$-$T$ and $n$-$T$ potentials are folded
with the deuteron ground state density.

We carry out the calculations for
total angular momentum $J$ up to 100, including $l=0$ and $2$ for the
$p$-$n$ relative angular momentum.
The radial discretization is
made with a spacing of $H = 0.2$ fm for $R$ and $h=0.5$ fm for $r$.
The matrix size is then given by  the  product of the number of angular 
momentum channel $N_J=4$, the number of radial grid points $N_R=250$ 
and $N_r=100$. This amounts to the total number of points 
$N_{\rm dim} = 10^5$. As mentioned in the previous section, we solve
this linear algebraic problem with the sparse complex matrix 
using  the Bi-conjugate-gradient method.

As for the grid spacing of the $p$-$n$ coordinate, we found
the spacing of 0.5 fm gives the accuracy of the deuteron binding energy 
better than 0.0001 MeV. The accurate binding energy 
by the $n$-$p$ potential employed is 2.2177 MeV, while the DVR calculations
with the grid spacing of 0.1 fm, 0.2 fm, and 0.5 fm give the same value
of 2.2177 MeV. 

\begin{figure*}
\includegraphics[width=120mm]{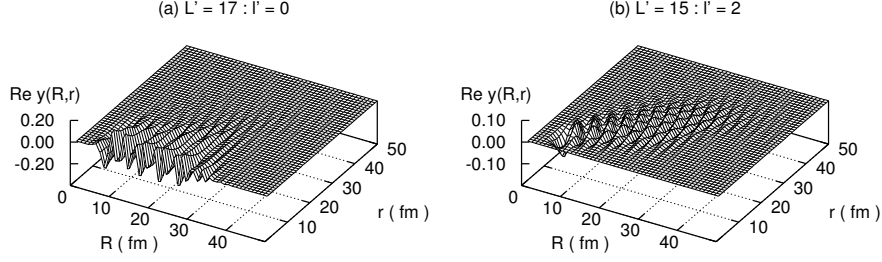}
\caption{\label{fig3}
(a) The real part of the radial wave function $y^J_{Ll_{\alpha},L'l'}(R,r)$ 
for $J=17$, $L'=17$, and $l'=0$ in the $d$-$^{58}$Ni scattering 
at $E_d=80$ MeV.
(b) The same as (a) but for $J=17$, $L'=15$, and $l'=2$.
}
\end{figure*}

In the present approach, we can explicitly obtain the wave function. 
In Fig.~\ref{fig3} (a), the real 
part of the radial wave function $y^J_{Ll_{\alpha},L'l'}(R,r)$ is displayed
for $J=L=L'=17$, and $l'=l_{\alpha}=0$. This includes
the elastic scattering wave as well as the breakup waves. 
In the small $r$ region   the wave function is   dominated 
by the elastic wave . It oscillates with the frequency close 
to $K_0$ in the coordinate of $R$. Its amplitude decreases 
as $R$ increases in the region $R > 25$ fm, because 
of the absorption by $-i\epsilon_{PT}(R)$. 
The wave function at large $r$ shows breakup 
 components  of the deuteron into  the  $p+n$ continuum state. 
The amplitude of the wave function also decreases at large $r$ 
 due to the absorption  by $-i\epsilon_{nC}(r)$. 
Fig.~\ref{fig3} (b) shows the real part of $y^J_{Ll_{\alpha},L'l'}(R,r)$
for $J =  17$, $L' = 15$, and $l'=2$. For this angular momentum
channel, only the breakup component is seen.

\begin{figure*}
\includegraphics[width=80mm]{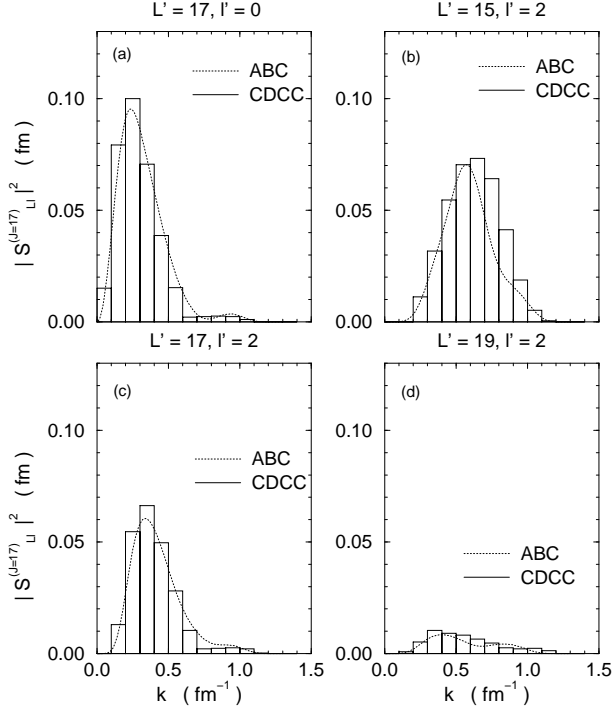}
\caption{\label{fig4}
(a) The squared moduli of the deuteron breakup $S$-matrix elements
$\vert S_{L=J l=0}^J\vert^2$. (b) $\vert S_{L=J-2 l=2}^J \vert^2$.
(c) $\vert S_{L=J l=2}^J\vert^2$.
(d) $\vert S_{L=J+2 l=2}^J\vert^2$.
In each panel, the dotted line and histogram denote 
the results of the ABC and CDCC methods, respectively.
}
\end{figure*}

We next show decomposition of the breakup cross section
$\sigma^J_{Ll}(k)$, which was defined in Eq.~(\ref{sgmJLl}). 
Following Ref.~\cite{yahiro1}, we show, in Fig.~\ref{fig4}, the
square moduli of the $S$-matrix, which are related to
$\sigma^J_{Ll}(k)$ by
$\sigma^J_{Ll}(k) = \frac{\pi (2J+1)}{K_0^2}
\vert S^{(J)}_{Ll}(k) \vert^2$ for the present case ($l_{\alpha}=0$).
The deuteron $s$-wave
breakup, $S^{(J)}_{L=J l=0}(k)$, are shown in  the  panel
(a) for $J$ = 17. Three kinds of $d$-wave breakup $S$-matrix elements
are shown in  the panels  (b) $S_{L=J-2 l=2}^{(J)}(k)$,
(c) $S_{L=J l=2}^{(J)}(k)$, and (d)
$S_{L=J+2 l=2}^{(J)}(k)$, respectively.
In each panel the dotted line and histogram represent 
the calculated results of the ABC and CDCC methods, respectively.
The global behavior of the smooth dotted lines agrees well
with that of the histogram
\footnote{There are small differences between the results obtained 
by the two methods, in particular,
$S_{L=15 l=2}^{J=17}$ around $0.7\ {\rm fm}^{-1} < k < 1.0\ {\rm fm}^{-1}$.
The agreement of the results between the two methods becomes much better if the
number of momentum bins is increased 
in the CDCC calculation \cite{matsumoto}.}.
We examined the sensitivity of the breakup cross sections
against the strength of the absorbing boundary potential.
We confirmed that the  calculated results 
are not sensitive to these parameters, 
as long as the condition of Eq.~(\ref{Wcond}) is satisfied. 

We show in Table \ref{tab:1} the total reaction cross section
$\sigma_R$ and the elastic breakup cross sections  $\sigma_{BU}$ .
The elastic breakup cross sections are decomposed according to the
$p$-$n$ relative angular momentum $l$.
While the total reaction cross section $\sigma_R$ 
and the $s$-wave breakup cross section $\sigma_{BU}^{(l=0)}$
calculated with the ABC method agree to those with the CDCC method, 
the $d$-wave breakup cross section
$\sigma_{BU}^{(l=2)}$ of the ABC method
is about 10 \% smaller than that of the CDCC method.
This difference in $\sigma_{BU}^{(l=2)}$ is attributed to the 
difference seen in Fig.~\ref{fig4}(b).

\begin{table}[t]
\caption{Reaction and breakup cross sections of $d$-$^{58}$Ni reaction
at $E_d=80$ MeV. The results of CDCC are taken from Ref.~\cite{yahiro1}.}
\label{tab:1}
\begin{center}
\begin{tabular}{lcccc} \hline\hline
     &  $\sigma_{BU}^{(l=0)}$ &  $\sigma_{BU}^{(l=2)}$ &  
$\sigma_{BU}$ &  $\sigma_{R}$ \\ \hline
ABC  & 38.28 mb & 77.21 mb & 115.49 mb & 1572.9 mb \\
CDCC & 37.77 mb & 88.59 mb & 126.36 mb & 1571.4 mb \\
\hline\hline
\end{tabular}
\end{center}
\end{table}

\begin{figure*}
\includegraphics[width=100mm]{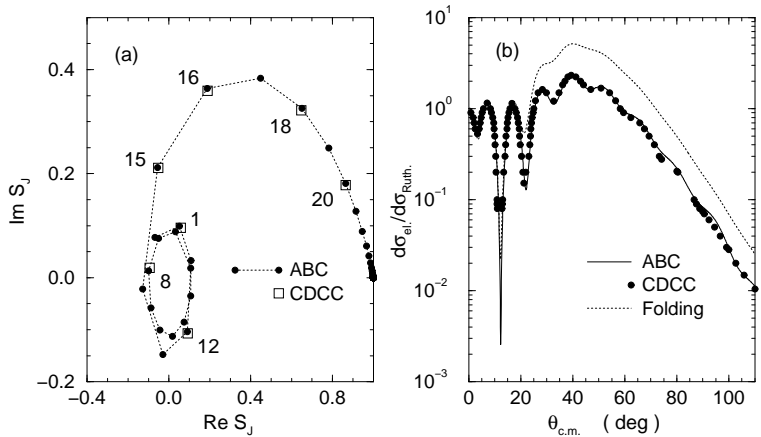}
\caption{\label{fig5}
The $S$-matrix (a) and differential cross section (b)
of $d$-$^{58}$Ni elastic scattering at $E_{d}$ = 80 MeV. 
Results of the ABC are compared with those of the CDCC \cite{yahiro}.
Results using a folding potential are also shown in (b).}
\end{figure*}

We next consider the angular distribution of the elastic scattering.
Fig.~\ref{fig5}~(a) shows the elastic scattering $S$-matrix elements $S_J$.
The closed circles with dotted line denotes 
$S_J$ calculated with the ABC methods. The open squares represent 
the corresponding CDCC results for $J$ = 1, 8, 12, 15, 16, 18, and 20.
Fig.~\ref{fig5}~(b) shows the elastic scattering differential cross section 
in Rutherford ratio. The differential cross
section calculated with the folding potential is also shown.
The difference between the ABC (CDCC) calculations and the folding
potential calculation shows the effect of the breakup process on
the elastic scattering. The results clearly show that the
ABC method gives  almost the same  results  as  the CDCC results.
Therefore, the ABC method can describe the influence of the breakup 
processes on the elastic scattering angular distribution.

\section{Summary}

The purpose of the present article is to show the usefulness of 
the ABC method to describe nuclear breakup reactions. 
The basic trick in the method is to place an absorbing potential
in the outer spatial region where complex three-body reactions do not
take place. The vanishing boundary condition is imposed outside 
the absorbing potential. If the absorbing potential works properly,
there exist only the outgoing waves at the edge of the interacting region 
where the absorbing potential is absent.
We can thus obtain scattering solution by solving the 
Schr\"odinger-like equation with a source term, without imposing
the scattering boundary condition.

In practical calculations, the equation is solved in the partial 
wave expansion, and the radial equation is discretized employing a 
uniform grid.
The problem is then converted into the linear algebraic equation of
large dimension with a complex, sparse, coefficient matrix.
We can efficiently solve the equation by employing the 
Bi-conjugate-gradient method with a pre-conditioning procedure.

To show the feasibility and the accuracy of the ABC method, we have 
applied the method to the $d$-$^{58}$Ni scattering as an example. 
For this scattering, detailed analyses with the CDCC method are available. 
We have found an excellent agreement between the results by the two 
methods, the ABC and the CDCC methods, for both breakup and elastic 
scattering. From this observation, we are confident that the 
physical contents of these two methods are the same, although the 
procedures in solving the three-body scattering problem look very different.

Let us compare the advantages and the limitations between the two methods.
One of the superior features of the ABC method is that it is easy to 
assess convergence of the calculated results. If we fix the radial grid 
points and the number of partial waves for the $n$-$C$ motion,
the quality of the absorbing potential determines
accuracy of the results. One can easily ascertain the reliability
by examining how sensitive the results are to changes
of the absorbing potential. The convergence check is more difficult in 
the CDCC method in which one must first prepare the discretized continuum 
wave functions. One must examine the delicate convergence as to the 
discretization of the continuum; the number of the momentum bins and 
the maximum momentum to be incorporated.

The principal drawback of the ABC method is its heavy computational 
cost. One must solve a linear algebraic equation of large dimension.
In the present example of deuteron reaction, one must solve the
equation of $10^5$ dimension for each angular momentum. 
For this problem, the computational time for each angular momentum 
is typically 20 minutes in a single alpha-21264 processor. This time 
multiplied by the number of angular momentum states gives the total 
computational time for the reaction at a fixed incident energy. Of course, 
the parallel computation would reduce the computational wall-time, by 
paralleling different angular momentum calculations.

In general, as the incident energy
gets higher, the calculation becomes more demanding. This is
because the finer grid spacing is required as the incident energy
becomes higher. Also the number of angular momentum states that
take part in the reaction increases. 
For reactions with heavy target nuclei, the Coulomb
breakup reactions are significant, in particular when the projectile
is bound extremely weak, such as the halo nuclei. The description
of such Coulomb breakup reactions will be a difficult problem for the
ABC method, because the coupling potential is so long-ranged that
one should employ a large critical radius $r_c$ to obtain reliable 
results. We must place the absorbing potential at a large radial 
distance, and must employ a large number of grid points. These 
difficulties are, however, not inherent in the ABC method, but
are suffered by the other methods as well.

Establishing the usefulness of the ABC approach for breakup reactions 
of a weakly bound projectile, we are now planning to apply the method 
to reactions of unstable nuclei. A preliminary result on the nuclear 
breakup reactions of a single neutron halo nucleus, $^{11}$Be, will 
be published in Ref.~\cite{yabana}. In that calculation, the number of 
the matrix size is about $2.4 \times 10^5$, not so much different 
from that in the deuteron reaction. The ABC method should, at least, be 
useful for breakup reactions of a single nucleon halo nucleus
induced by the nuclear force.
Applicability of the ABC method
depends on the availability of the computational facilities. We expect 
that the ABC method will be more popular and useful as the parallel 
computation becomes more prevalent.

The ABC method is also useful in any circumstances where the scattering
boundary condition comes in.
Some applications have been done and are in progress concerning
resonances of a few-body system and response in the continuum,
especially, of deformed systems \cite{nakatsukasa,nakatsukasa1,nakatsukasa2}.

\begin{acknowledgments}
One of the authors (T.N.) acknowledges the Grand-in-Aid
for Scientific Research (No. 14740146) from the Japan Society for
the Promotion of Science.
\end{acknowledgments}

\end{document}